\documentclass[12pt,letterpaper,leqno,doublespacing]{article}
\usepackage[top=0.9in,bottom=0.9in,left=0.9in,right=0.9in]{geometry}
\usepackage{graphicx}
\usepackage{upgreek}
\usepackage{amsmath}
\usepackage{amsfonts}
\usepackage{setspace}
\usepackage{relsize}
\usepackage{bm}
\usepackage[title]{appendix} 
\usepackage{multirow,array}
\usepackage{float}
\usepackage[table]{xcolor} 
\usepackage{authblk}
\usepackage{rotating}
\usepackage{lscape}   
\usepackage{titlepic}  
\usepackage{adjustbox} 
\usepackage{fancyhdr}
\usepackage{booktabs,tabularx}
\usepackage{comment}

\usepackage{setspace,epstopdf,amsmath,amsfonts,amssymb,amsthm}
\usepackage{marginnote,datetime,enumitem,rotating,fancyvrb}
\usepackage[draft,hidelinks]{hyperref}
\usepackage{float}
\usdate
\hypersetup{  
	colorlinks=true,
	linkcolor=black,
	filecolor=black,      
	urlcolor=black,
	citecolor= black,
	bookmarks=true,
}
   
\DeclareMathOperator*{\argmin}{arg\,min}

\usepackage{cleveref} 
\usepackage{subfig}

\usepackage{setspace}
\usepackage{footmisc}

\usepackage[labelfont={bf,small},textfont={bf,small},format=hang]{caption}
\makeatletter
\renewcommand\@seccntformat[1]{\csname the#1\endcsname.\quad}
\makeatother
\usepackage{natbib}
\setcitestyle{authoryear}

\author{Muzi Chen, Yuhang Wang, Boyao Wu, and Difang Huang}

\begin{document}
	
	
	
	\title{\textbf{Dynamic Analyses of Contagion Risk and Module Evolution on the SSE A-Shares Market Based on Minimum Information Entropy}}
	\date{\vspace{-5ex}}
	\maketitle

\begin{abstract}
	The interactive effect is significant in the Chinese stock market, exacerbating the abnormal market volatilities and risk contagion. Based on daily stock returns in the Shanghai Stock Exchange (SSE) A-shares, this paper divides the period between 2005 and 2018 into eight bull and bear market stages to investigate interactive patterns in the Chinese financial market. We employ the LASSO method to construct the stock network and further use the Map Equation method to analyze the evolution of modules in the SSE A-shares market. Empirical results show: (1) The connected effect is more significant in bear markets than bull markets; (2) A system module can be found in the network during the first four stages, and the industry aggregation effect leads to module differentiation in the last four stages; (3) Some stocks have leading effects on others throughout eight periods, and medium- and small-cap stocks with poor financial conditions are more likely to become risk sources, especially in bear markets. Our conclusions are beneficial to improving investment strategies and making regulatory policies.
\end{abstract}

\textit{Keywords}: Map Equation; minimum information entropy theory; module detection; LASSO method; industry aggregation; network analysis
\bigskip

\clearpage

\newpage
\section{Introduction}

From June 2014 to June 2015, the Shanghai Stock Exchange (SSE) A-shares index increased by $ 57\% $. The market then experienced three large-scale collapses during the following half year from June 2015 to January 2016, where the index decreased by $ 49\% $ \citep{darby2021institutional}. During these events, the SSE A-shares market plummeted with high volatility and lost about 36 trillion RMB. Such abnormal fluctuations in the stock market were also accompanied by the highly synergistic effect of the rise and fall of the stock market, further increasing the stock market's volatility \citep{ding2020valuation}.

Although the China Securities Regulatory Commission has conducted a comprehensive reform since 2005, there are repeated abnormal fluctuations over the transition period between the bull and bear market in the SSE A-shares market partly due to the immaturity of the Chinese stock market in the aspects of traders, trading system, market system, and regulatory system \citep{zhang2021turnover}. During expansions (bull markets), stocks in the A-shares market display blow-out increases, accumulating bubbles and financial risks. During recessions (bear markets), the fire sale trading triggers the declines in stock liquidity and the spreads the financial risk throughout the entire financial system \citep{liu2020a}.

As the second-largest economy globally, China devotes itself to integrating into the global finance market. Specifically, a series of policy measures, including establishing the Renminbi Qualified Foreign Institutional Investor (RQFII) program, opening the Shanghai-Hong Kong Stock Connect, and continuously raising quotas in the Qualified Foreign Institutional Investor (QFII) program and the Qualified Domestic Institutional Investor (QDII) scheme, are adopted to significantly strengthen the connections between China and the rest of the world \citep{jiang2019the}. The evolution of the SSE A-shares market over good and bad periods is related to the reform of the Chinese stock market and has profound influences on the international capital and cross-border spillover risk \citep{Wu2019,Yu2020}.

Traditional econometric measurements, including the Pearson correlations and Granger causalities, qualify the pair-wise relationship of two concerned stocks in a financial market without considering the potential influence from rest ones in the same system \citep{billio2012econometric,engle2002dynamic,engle2012volatility}. Although the multivariate regression model can characterize interactions between equities from a systemic way, this framework may fail to effectively fit the real data due to the high-dimensional problem that limited observations are used to estimate a significant number of parameters reflecting relationships of stocks \citep{acemoglu2015systemic,acharya2012capital,diebold2008measuring,diebold2011on}. To overcome these above issues, we use the Least Absolute Shrinkage and Selection Operator (LASSO) method to model the SSE A-shares market returns and measure statistically significant connections between the equities system, and shrink those insignificant ones into zeros \citep{Wu2019,yan2020development}. More importantly, we also use the Map Equation method to conduct a dynamic analysis of financial contagion patterns. Our empirical results show that the SSE A-shares market's contagion pattern reveals the industry differentiation after 2014, and compared with large-cap stocks, medium- and small-cap stocks react to financial risks more distinctly and function as financial contagion channels.



The interactions between stocks and contagion risks are essential to understand the stock market fluctuations and global financial crisis \citep{girardi2013systemic,hautsch2015financial,kyrtsou2020exploitation}. The classical econometric methods rely on pair-wise measurements including the Pearson correlations to describe the relationships within the network system \citep{adeloye2015an,adeloye2015global,papana2013simulation,rudan2015prevalence}. \citet{naoui2010a} use the DCC-GARCH model to study the pair-wise relationships of stock index returns of different regions over the subprime mortgage crisis and find that the United States is an essential source of contagion during this crisis. Selecting three fields--industry, banking, and public utilities--as research objects, \citet{grout2016stock} show that industrial market risks increase during the crisis. \citet{bernal2014assessing} introduce the CoVaR method to measure the relationships among stock returns in the banking, insurance, and other financial sectors during the financial crisis. \citet{das2019the} propose a mixed-frequency based regression approach, derived from functional data theories, to analyze the influence of global crises on stock market correlations between G7 countries.

Classical econometric approaches mainly focus on the direct relationship between two financial agents but fail to reflect potential influences from the direct connection's complex systems \citep{battiston2012liaisons,bisias2012a,huang2009a,kritzman2011principal}. However, such underlying interactions can be well revealed under the networking framework by investigating financial networks' topological properties and statistical characteristics. \citet{liu2012a} use five years of stock index data from 67 countries and use Pearson correlations to generate a complex network. \citet{gong2019measuring} employ the transfer entropy method to analyze interactions between national stock markets and discover that countries affected by the crisis become closer to each other and the total network connectedness rises during the crisis. \citet{chen2020correlation} use complex network theories to measure systemic risks in the stock market and develop dynamic topological indicators to analyze financial contagion and qualify the magnitude of systemic risks. Constructing undirected and directed volatility networks of the global stock market, \citet{lee2019global} apply machine learning methods to study network indicators for establishing an international financial portfolio management approach. \citet{liu2017interbank} base on 6,600 banks' decision rules and behaviors reflected in quarterly balance sheets to construct interbank networks and further examine how financial shocks spread through financial contagion. \citet{kumar2012correlation} apply random matrix theories to study topological properties of the network consisting of 20 nations and analyze communities in the generated network under different thresholds. \citet{li2017global} discuss the relationships between listed energy companies and their shareholders under the networking framework. Empirical studies show that most energy investment is concentrated in a few countries, and some islands or regions play irreplaceable roles in tax avoidance. \citet{paltalidis2015transmission} employ the maximum entropy method to study the systemic risk and analyze the vulnerability of the Euro area's financial network.

Using networking methodologies to study stock markets may suffer from the high-dimensional problem such that traditional estimations are not consistent. The number of stocks ($ N $) in a market is comparable to that of observations ($ T $) over a specific period (e.g., the bull and bear market), and hence the size of unknown parameters ($ O(N^2) $) is comparable to that of data ($ O(NT) $). The LASSO method provides a promising solution to alleviate the high-dimension problem when building financial networks. \citet{xu2019interconnectedness} utilize the LASSO-CoVaR model to construct a financial network for the Chinese stock market between 2010 and 2017 and analyze financial institutions' status and role in crises. Using the data on the subprime mortgage crisis, \citet{demirer2018estimating} adopt the LASSO-VAR method to analyze the static and dynamic connectedness in the global system.

Applying the LASSO method, we generate the stock networks and community structures to analyze the evolution of financial systems. We further introduce the Map Equation method to study the dynamic changes of the SSE A-shares market and its differentiation in industries. The Map Equation method is based on information theories and further improved in subsequent studies \citep{rosvall2007an, rosvall2008maps, rosvall2009the}, widely used in biological, information, and social networks \citep{kim2011map}. This method utilizes the probability flow of random walks on a network as a proxy for information flows in the entire system to decompose the network into different modules by compressing the probability flow description. The Map Equation approach is also adopted to investigate risk transmission in financial settings, which makes it possible to analyze the overnight market risk path of commercial banks \citep{bech2015mapping}, the centrality of financial network institutions and measurements on systemic risks \citep{chan-lau2018systemic}, and the financial integration of banks in developed regions before the subprime crisis \citep{garratt2014the}.


\section{Methodologies}

\subsection{LASSO Estimation and Network Construction}

Traditional econometric methods use pair-wise measurements, such as Pearson correlations and Granger causalities, to qualify interactions between stocks. However, these approaches only measure the direct relationship between the two stocks without considering the potential influence from the rest stocks through a systematic perspective. When using Pearson correlations, the correlation between stock A and B may indirectly derive from stock C, which is highly correlated with stock A and B separately. Granger causalities are also inappropriate for describing sophisticated linkages in financial markets due to theoretical reasons. For any given
stock pair, the white noise assumption in the Granger causality test implies that there are no connections between the concerned two stocks and the rest. In other words, these two stocks are presumed to be isolated from the entire system, and this may contradict the networking structure in financial markets. Although classic multivariate regression models can overcome the above two drawbacks, they fail to fit the data in the high-dimensional situation that the number of stocks ($ N $) is proportional to that of observations ($ T $) (i.e., $ N = O(T) $). As a promising solution, the LASSO method chooses an absolute value function as the penalty term to screen significant variables and shrinks insignificant ones into zeros, which can solve the fitting problem in high-dimensional cases.

We consider a multivariate linear regression model to reveal relationships in the stock market from a systemic way. For the stock $ i $, the model is
\begin{equation}
	r_{it} = r_{1t} \beta_{i1} + r_{2t} \beta_{i2} + \cdots + r_{i-1,t} \beta_{i,i-1} + r_{i+1,t} \beta_{i,i+1} + \cdots + r_{Nt} \beta_{iN} + \varepsilon_{it},
	\label{Eq Multivariate Linear Regression Model for Stock i}
\end{equation}
where $ r_{it} = \ln P_{it} -  \ln P_{i,t-1} $ is the log return of stock $ i $, $ P_{it} $ is the stock price of $ i $ at time $ t $, $ N $ is the number of stocks and $ \varepsilon_{it} $ is the error term. Since unknown parameters $ \bm{\beta}_{-i} = \{ \beta_{ij}, j = 1, \cdots, i-1, i+1, \cdots, N \} $ qualify the stock relationships in the financial market, the LASSO method is adopted to estimate those statistically significant parameters and shrink those insignificant ones into zeros. The LASSO estimate is
\begin{equation}
	\widehat{\bm{\beta}}_{-i} = \argmin_{\bm{\beta}_{-i}} \left\lbrace \frac{1}{2T} \sum_{t=1}^{T} \left( r_{it} - \sum_{j \neq i} r_{jt} \beta_{ij} \right)^2 + \lambda \sum_{j \neq i} \big| \beta_{ij} \big| \right\rbrace,
	\label{Eq LASSO Estimation for Stock i}
\end{equation}
where $ T $ is the number of observations and $ \lambda $ is the tuning parameter preset by the cross-validation method.\footnote{This paper uses the glmnet package in R to obtain LASSO estimations.} Repeat the above procedure for all stocks. Then, the adjacency matrix $ \bm{A} = \{a_{ij}\} $ for the financial market is define as
\begin{equation}
	a_{ij} = \left\lbrace \begin{array}{cc}
		1, & \mathrm{if} \ \widehat{\beta}_{ji} \neq 0;   \\
		0, & \mathrm{if} \ \widehat{\beta}_{ji}   =  0.
	\end{array} \right.
	\label{Eq Adjacency Matrix}
\end{equation}
Equation \eqref{Eq Adjacency Matrix} suggests that a direct link from stock $ i $ to $ j $ exists if and only if its corresponding LASSO estimate $ \widehat{\beta}_{ji} $ is non-zero.


\subsection{Module Detections Based on Information Entropy Methods}

This paper introduces the Map Equation method to detect modules in the SSE A marker network over different periods and further explores the evolution of modules. The Map Equation algorithm (InfoMap algorithm), initially proposed by \citet{rosvall2007an}, is based on a formula to evaluate the effectiveness of a module structure in describing the path of a random walker around the network. The random walk is used to simulate the information (risk) transmission in the entire system (the stock market). Based on the Huffman coding, the Map Equation algorithm adopts a two-level code to describe the random walk path: the high-level codes distinguish modules in the network (i.e., index codes), and the low-level codes represent node names that are unique in the same group (i.e., module codes). The Map Equation algorithm aims to discover an information (risk) map that gets rid of unnecessary details by minimizing the amount of information needed to describe the random walk path and highlights critical modules where nodes in the same group develop stronger interior relationships than outside nodes.

Since the real relationships between stocks are not observable, we first base on the discovered network structure in Eq. \eqref{Eq Adjacency Matrix} to simulate the real one and calculate the visit frequencies of a random walker traveling in the system. Similar to the used method in \cite{garratt2014the}, we convert the adjacency matrix $ \bm{A} $ defined in Eq. \eqref{Eq Adjacency Matrix} to the Markov transition probability matrix $ \varPi $ to depict the random walk path. Since $ \bm{A} $ may be non-symmetric, we consider the follow systemic matrix
\begin{eqnarray}
\bm{V} = \{v_{ij}\} = \left[ \begin{array}{cc}
\bm{0}_{N}   &   \bm{A}   \\
\bm{A}^\top  &   \bm{0}_N   \\
\end{array} \right] \in \mathbb{R}^{2N \times 2N},
\end{eqnarray}
where $ \bm{0}_N \in \mathbb{R}^{N \times N} $ is a matrix with zeros. Then, the Markov transition probability matrix is defined as
\begin{eqnarray}
\varPi = \left[ \pi_{ij} \right]_{2N \times 2N} = \left[ \frac{v_{ij}}{\sum_{k=1}^{2N} v_{kj}} \right]_{2N \times 2N}.
\label{Eq Markov Transition Probability Matrix}
\end{eqnarray}
Let $ p_i $ be the visit frequency of a random walker to the node $ i $ . Mathematically, we can calculate $ p_i $ by considering the dominant eigenvector of the Markov transition probability matrix
\begin{eqnarray}
\bm{P} = \varPi \bm{P},
\end{eqnarray}
where $ \bm{P} = \left( p_1, \cdots, p_{2N} \right)^\top $.

Given by the visit frequencies $ \bm{P} $ and a module structure $ \bm{M} $ with $ m $ modules, the exiting frequency of the traveler from module $ \alpha $ is given by
\begin{equation}
q_{\alpha \curvearrowright} = \sum_{i \in module \ \alpha} \sum_{j \notin module \ \alpha} \pi_{ij} p_i,
\end{equation}
and the exit frequency of the travel from \emph{any} module is given by
\begin{eqnarray}
q_{\curvearrowright} = \sum_{\alpha = 1}^{m} q_{i \curvearrowright}.
\end{eqnarray}
Moreover, the frequency that the random walker use module $ \alpha $'s codes is given by
\begin{equation}
p_{\circlearrowleft}^{\alpha} = q_{\alpha \curvearrowright} + \sum_{i \in module \ \alpha} p_i.
\end{equation}

The probabilities $ p_{\circlearrowleft}^{\alpha} $ and $ q_{\alpha \curvearrowright} $ measure the frequency of using module codes. Next, we need to know the costs to access these codes. According to the Shannon's coding theorem, for a random variable $ z $ having $ n $ states with probabilities $ p_k $, the average length of the code word cannot be less than the entropy of $ z $, defined by
\begin{equation}
H(z) = - \sum_{k=1}^{n} p_k \log (p_k).
\label{Eq Information Entropy}
\end{equation}
Then, the minimum average length for index codes and module codes are given by
\begin{equation}
H(Q) = - \sum_{\alpha = 1}^{m} \frac{q_{\alpha \curvearrowright}}{q_{\curvearrowright}} \log \left(\frac{q_{\alpha \curvearrowright}}{q_{\curvearrowright}}\right)
\end{equation}
and
\begin{equation}
H(P^\alpha) = - \frac{q_{\alpha \curvearrowright}}{p_{\circlearrowleft}} \log \left(\frac{q_{\alpha \curvearrowright}}{p_{\circlearrowleft}}\right) - \sum_{i \in module \ \alpha} \frac{p_i}{p_{\circlearrowleft}^{\alpha}} \log \left(\frac{p_i}{p_{\circlearrowleft}^{\alpha}}\right),
\end{equation}
where $ H(\cdot) $ is the entropy function defined in Eq. \eqref{Eq Information Entropy}, $ H(Q) $ and $ H(P^\alpha) $ reflect the encoding effectiveness of the entire module structure $ M $ and the specific module $ \alpha $, respectively.

Consequently, the minimum average length of the random-walk path under the given module structure $ M $ is given by
\begin{equation}
L(M) = q_{\curvearrowright} H(Q) + \sum_{\alpha=1}^{m} p_{\circlearrowleft}^{\alpha} H(P^\alpha).
\end{equation}
Here, $ L(M) $ is the weighted sum of two information entropy parts. The one is the average code length of the module name (index codes), and the other one is the average code length of node names in different modules (module codes). The weights are proportions of average code lengths of module and node names. When minimizing the $ L(M) $, the network achieves the minimum entropy, and the corresponding module division is stable. The map equation algorithm uses these criteria to compare different module divisions in practice and select the optimal one.


\subsection{Network topological indicators}
\label{Section Network topological indicators}

Based on the detected module structures, we further investigate the topological properties of generated networks to study the contagion effect in the SSE A market. 
\begin{description}

\item[Average shortest path length] $ L = \frac{2}{N(N-1)} \sum_{j \neq i} d_{ij} $, where $ i $ and $ j $ are two stocks (nodes) in the SSE A network and $ d_{ij} $ is the shortest path between nodes $ i $ and $ j $. A smaller length means faster information or risk transmission in the network.

\item[Clustering coefficient] $ C_i = \frac{2 n_i}{k_i (k_i - 1)} $, where $ k_i $ is the number of nodes directly connecting to node $ i $ and $ n_i $ is the number of edges between $ k_i $ neighbours of node $ i $. A higher value implies better network connectivity.

\item[Network diameter] $ \mathrm{Diameter} = \max_{1 \leq i,j \leq N} d_{ij} $. A smaller value implies faster information or risk transmission speed.

\item[Network density] $ \mathrm{Density} = \frac{\sum_{i,j} a_{ij}}{N (N-1)} $, where $ a_{ij} $ is defined in Equation \eqref{Eq Adjacency Matrix}. A higher density implies closer relationships between nodes.

\item[Relative degree centrality] $ C_{RD} (i) = \frac{k_i}{N-1} $. The high relative degree centrality implies the important influence from the corresponding node on the network. 

\item[Relative betweenness centrality] $ C_{RB} (i) = \frac{2}{(N-1)(N-2)} \sum_{j < k} \frac{g_{jk} (i)}{g_{jk}} $, where $ g_{jk} (i) $ is the number of shortest paths connecting nodes $ j $ and $ k $ and passing through node $ i $. This indicator measures the ``bridge'' role of node $ i $ in the network.

\item[Relative closeness centrality] $ C_{RC} (i) = \frac{N-1}{\sum_{j=1}^{N} d_{ij}} $ that measures how close node $ i $ is to all other nodes in the network. The high value of relative closeness centrality implies close connections between node $ i $ and other nodes.

\item[Degree centralisation] $ C_D = \frac{\sum_{i=1}^{N} \left( C_{RD} (\max) - C_{RD}(i) \right)}{\max \left[ \sum_{i=1}^{N} \left( C_{RD} (\max) - C_{RD}(i) \right) \right]} $, where the numerator is the sum of differences between the maximum degree centrality $ C_{RD} (\max) = \max_{1 \leq i \leq N} C_{RD}(i) $ and the degree centrality of each node $ C_{RD} (i) $ and the denominator is the maximum value of numerator in theories. This indicator describes the centrality of the whole network.

\item[Betweenness centralisation] $ C_B = \frac{1}{N-1} \sum_{i=1}^{N} \left( C_{RB} (\max) - C_{RB}(i) \right) $, where the numerator theoretically represents the sum of the difference between the maximum intermediate centrality and the intermediate centrality of each node, and the denominator represents the maximum of the sum of the differences. This indicator describes the degree to which the network relies excessively on a node to transfer relations.

\item[Closeness centralisation] $ C_C = \frac{2(N-3)}{(N-1)(N-2)} \sum_{i=1}^{N} \left( C_{RC} (\max) - C_{RC}(i) \right) $ that describes the centralized trend in the network.

\end{description}


\section{Empirical Results}

\subsection{Stock Market Data}

This paper uses the SSE A-shares market's weekly closing prices from 2005 to 2018 and divides the entire period into eight stages according to bull and bear markets.

\begin{table}[H]
	\centering
	\caption{The period division from 2005 to 2018 for the SSE A-shares market according to bull and bear markets.}
	\label{Tab Period Division from 2005 to 2018}
	\footnotesize
	\begin{tabular}{llccl}
		\hline
		\multicolumn{1}{c}{Eight stages} & \multicolumn{1}{c}{Code} & Start time & End time & Reasons                                                                                             \\ \hline
		Stage 1                          & BULL1                    & 2005-06       & 2007-10  & Reforms in the market and other capital dividends.                                                  \\
		Stage 2                          & BEAR1                    & 2007-10       & 2008-10  & The subprime mortgage crisis and other external factors.                                            \\
		Stage 3                          & BULL2                    & 2008-10       & 2009-07  & Rescue policies from the government.                                                                \\
		Stage 4                          & BEAR2                    & 2009-07       & 2014-03  & Combined influences from the crisis and stimulus policies  \\
		& & & &  result in market ups and downs. \\
		Stage 5                          & BULL3                    & 2014-03       & 2015-06  & Deepen reforms of state-owned enterprises and arising \\
		& & & &  financial leverages increase the market.      \\
		Stage 6                          & BEAR3                    & 2015-06       & 2016-01  & Deleveraging and other factors cause the collapse of \\
		& & & &  the stock market.                              \\
		Stage 7                          & BULL4                    & 2016-01       & 2018-01  & The slight rise in the market due to factors like stable leverage \\
		& & & &  and financial stimulus.    \\
		Stage 8                          & BEAR4                    & 2018-01       & 2018-12  & Overseas factors like the trade war leads to the fluctuating \\
		& & & &  decline in the stock market.        \\ \hline
	\end{tabular}
	
\end{table}

As shown in Table~\ref{Tab Period Division from 2005 to 2018}, stages 1 and 5 are typical rapid-rising bull markets, and stages 2 and 6 are steeply declining bear markets. By contrast, stages 3 and 7 witness moderate increases in the market, and the market experiences a fluctuating decline in stages 4 and 8. Overall, the surge and plummet can be found in the bull market (stages 1 and 5) and the bear market (stages 2 and 6), respectively, whereas the stock market consistently fluctuates in stages 3, 4, 7, and 8. Based on the above division, we aim to utilize the information entropy to discover the evolution of modules in the SSE A-shares market in expansions and recessions.

The SSE A-shares backward closing prices from 2005 to 2018 are downloaded from the wind database. Given the long period of the research data, some stocks are out of the discussion in this paper because of the following reasons.
\begin{enumerate}
	\item Missing values. Until 18 December 2019, 1,547 stocks are traded on the SSE A-shares market. Due to the late listing and suspension of trading, some of these stocks are removed in advance to compare stock market networks in different stages.
	
	\item Stocks prefixed with ``ST'' or ``*ST''. Designated as Special Treatment (ST) by the stock exchanges for warning investors, these stocks face delisting risks and show distinct patterns from normal stocks.
	
	\item Stocks whose returns maintain at zeros over a long period for the long-term suspension or other reasons are excluded in this paper to prevent misleading information.
\end{enumerate}
As a result, 716 stocks are selected to construct the SSE A-shares network. For the stock $ i $, its log return is calculated by
\begin{equation}
	r_{it} = \ln P_{it} -  \ln P_{i,t-1},
	\label{Eq Log Returns}
\end{equation}
where $ P_{it} $ is the price of stock $ i $ at time $ t $.


\subsection{Module Analysis of the SSE A-shares Network Based on the Minimum Entropy}

\subsubsection{Network Module Visualisation}
To analyze the characteristics of modules in the SSE A-shares market, we employ the Map Equation method to detect modules in the financial system in different periods and present visualization results in Figure~\ref{Fig Module Visualisation}.\footnote{Figures~\ref{Fig Full Size 1} and \ref{Fig Full Size 2} in the Appendix provide the full-size version of Figure~\ref{Fig Module Visualisation}.} Each module is represented by a node whose area is proportional to its information flow, reflecting its status in the network. The information flow of a module includes two parts: the information flow out of the module, which is proportional to its boundary thickness and can represent the probability of risk passing from the module to other modules; and the information flow staying in the module, which is proportional to the area of the inner circle area that can indicate the probability that the risk remains in the module. The connection thickness among modules suggests the probability of risk transmission in different sectors. The thicker the connection, the greater the contagion probability. Besides, the arrow indicates the direction of the risk of contagion.

\begin{figure}[H]
	\includegraphics[width = \columnwidth]{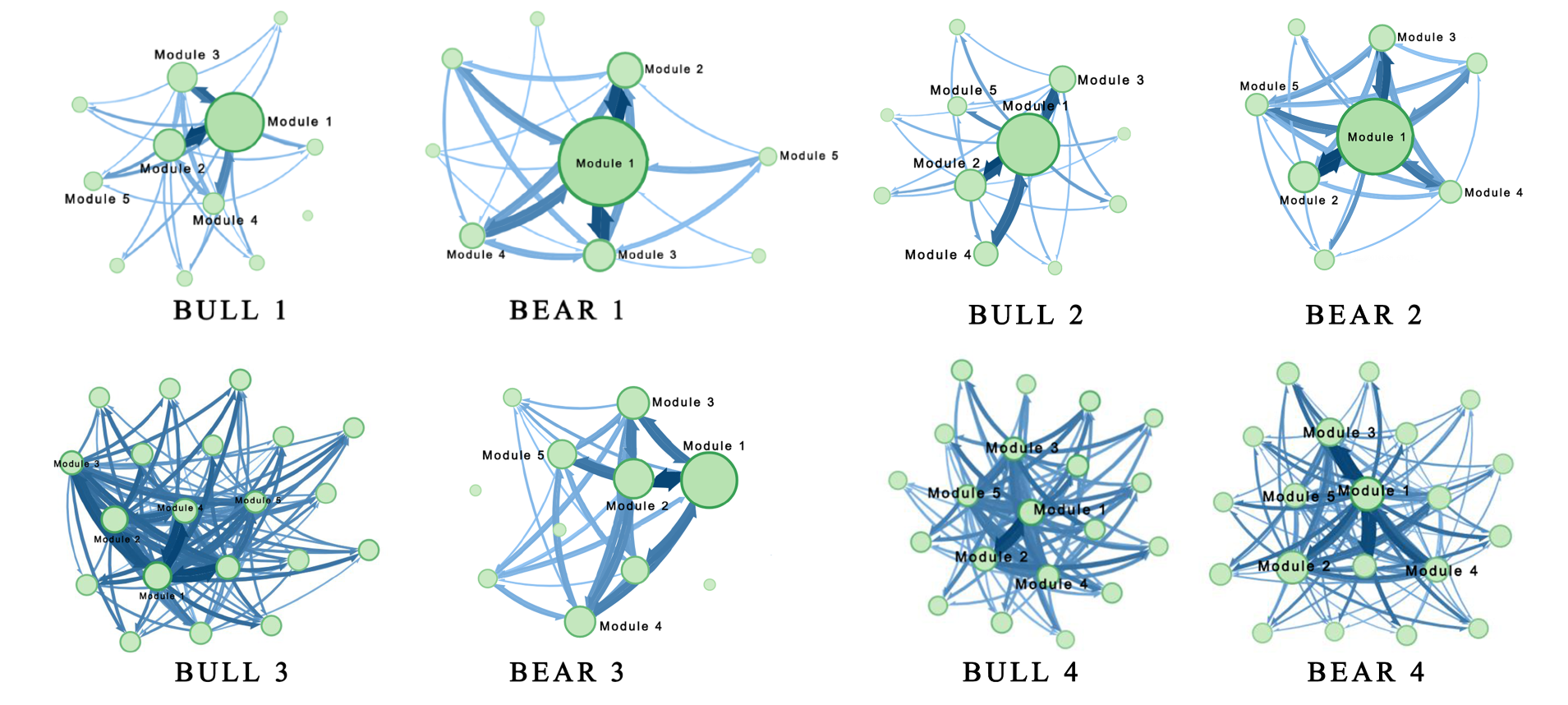}
	\caption{Module divisions in eight periods based on the Map Equation algorithm. Nodes represent modules, directed links indicate directions of the information flow and the thickness of links demonstrates the transmission probability of risk between different modules. The node size is proportional to the information flow in a module that includes the information flow out of and within the module. Specifically, the node's boundary thickness is proportional to the information flow out of the module, corresponding to the probability that risks transmit to other modules. The interior area of a node is proportional to the module's information flow, corresponding to the probability that risks staying in the module.
	}
	\label{Fig Module Visualisation}
\end{figure}

As shown in Figure~\ref{Fig Module Visualisation}, the module numbers in bear markets are generally smaller than those in bull markets, and more stocks belong to the same module in bear periods, indicating closer interior connections and higher internal contagion in modules. The number of modules in the stock market gradually increases, implying the continuous development of the capital market leads to the differentiation in the patterns of stock returns and the risk of contagion.


\subsubsection{Module Analysis Based on Map Equation Algorithm}
Based on the Map Equation method and the minimum information entropy, we discuss the module divisions in eight periods and further analyze the risk spillover among modules.

Table~\ref{Tab Node and Link Numbers of Top Five Modules} compares the top five modules with the highest proportion of information in bull and bear periods. The first column of Table~\ref{Tab Node and Link Numbers of Top Five Modules} shows the modules ranked by information flow from high to low, the second column represents the number of stocks (nodes) involved in the module, and the third column represents the number of information transmission linkages (links) within the module. Table~\ref{Tab Node and Link Numbers of Top Five Modules} suggests significant structural differences exist between the first four stages and the latter four stages. In the first four stages, the proportions of stock and link numbers within the largest module account for more than $ 70\% $ of the entire SSE A-shares network, whereas the proportions in other modules are low. In other words, the SSE A-shares market does not show significant differentiation in the first four stages: the largest module having the highest information flow largely represents the entire network, and more than $ 70\% $ of the stocks belong to this module. However, the SSE A-shares market experiences significant structural differentiation in the last four stages. Specifically, the largest five modules in BULL3, BULL4, and BEAR4 share comparable sizes in node and link numbers. Compared with the other three stages, the BEAR3 period is closer to the first four stages because the rapid deleveraging effect in this stage leads to abnormal fluctuations and increases the system's connected effect.

\begin{table}[H]
	\centering
	\caption{Node and link numbers of the top five modules with the highest proportion of information in eight periods.}
	\label{Tab Node and Link Numbers of Top Five Modules}
	\footnotesize
	\begin{tabular}{ccc|ccc}
		\hline
		\multicolumn{3}{c|}{BULL1}                & \multicolumn{3}{c}{BEAR1}                \\ \hline
		Module name & Node numbers & Link numbers & Module name & Node numbers & Link numbers \\ \hline
		M1    & 488          & 9637         & M1    & 608          & 10887        \\
		M2    & 86           & 688          & M2    & 44           & 307          \\
		M3    & 61           & 468          & M3    & 33           & 224          \\
		M4    & 26           & 161          & M4    & 19           & 49           \\
		M5    & 11           & 33           & M5    & 4            & 8            \\ \hline
		\multicolumn{3}{c|}{BULL2}                & \multicolumn{3}{c}{BEAR2}                \\ \hline
		Module name & Node numbers & Link numbers & Module name & Node numbers & Link numbers \\ \hline
		M1    & 528          & 11755        & M1    & 612          & 14331        \\
		M2    & 47           & 329          & M2    & 52           & 802          \\
		M3    & 51           & 797          & M3    & 15           & 50           \\
		M4    & 43           & 533          & M4    & 8            & 18           \\
		M5    & 20           & 166          & M5    & 10           & 23           \\ \hline
		\multicolumn{3}{c|}{BULL3}                & \multicolumn{3}{c}{BEAR3}                \\ \hline
		Module name & Node numbers & Link numbers & Module name & Node numbers & Link numbers \\ \hline
		M1    & 53           & 231          & M1    & 306          & 3750         \\
		M2    & 47           & 264          & M2    & 119          & 1147         \\
		M3    & 31           & 119          & M3    & 70           & 500          \\
		M4    & 27           & 119          & M4    & 72           & 503          \\
		M5    & 27           & 123          & M5    & 69           & 504          \\ \hline
		\multicolumn{3}{c|}{BULL4}                & \multicolumn{3}{c}{BEAR4}                \\ \hline
		Module name & Node numbers & Link numbers & Module name & Node numbers & Link numbers \\ \hline
		M1    & 40           & 256          & M1    & 95           & 668          \\
		M2    & 33           & 211          & M2    & 68           & 476          \\
		M3    & 30           & 122          & M3    & 52           & 328          \\
		M4    & 39           & 201          & M4    & 22           & 61           \\
		M5    & 29           & 108          & M5    & 36           & 126          \\ \hline
	\end{tabular}
	
\end{table}

We use the information flow to illustrate the spillover effect of risks within and between modules in Table~\ref{Tab Information Flows Within and Between Modules}: The first column represents the modules ranked by information flow from high to low; the second column represents the proportions of information within modules to the total information; the third (fourth) column represents the proportion of information flowing in (out) each module. Table~\ref{Tab Information Flows Within and Between Modules} presents similar results as Table~\ref{Tab Node and Link Numbers of Top Five Modules}. In the first four stages, the information in the largest module accounts for at least $ 70\% $ of the entire market, indicating that most stocks can be grouped and insignificant differentiation exists in the stock market. Conversely, in the last four stages, except for the BEAR3 stage, the most extensive module merely contains approximately $ 10\% $ information, suggesting significant differentiation in the stock market. Such differentiation derives from industry agglomeration that stocks from the same industry are more likely to form modules.\footnote{See details in Tables~\ref{Tab Industry Categories of Stocks in BULL4} and \ref{Tab Industry Categories of Stocks in BEAR4}.} Moreover, the BEAR3 can be viewed as a transition period between the first and last four stages since the information proportion of M1 is about $ 50\% $. During this period, given the excessive accumulation of pre-leverage and the speed of the deleveraging process, abnormal fluctuations occurred in the stock market, and the strengthened connected effect in the system further increases the abnormal volatility of the stock market, resulting in a more significant loss in the financial system.

\begin{table}[H]
\centering
\caption{Information flows within and between modules over eight periods.}
\label{Tab Information Flows Within and Between Modules}
\footnotesize
\begin{tabular}{cccc|cccc}
	\hline
	\multicolumn{4}{c|}{BULL1}               & \multicolumn{4}{c}{BEAR1}              \\ \hline
	Module   & Within  & Flow in & Flow out & Module   & Within  & Flow in & Flow out \\
	Name     & Modules & Modules & Modules  & Name     & Modules & Modules & Modules  \\ \hline
	M1 & 66.13\% & 8.58\%  & 10.82\%  & M1 & 85.87\% & 5.29\%  & 5.94\%   \\
	M2 & 13.32\% & 6.00\%  & 5.41\%   & M2 & 6.80\%  & 3.14\%  & 2.75\%   \\
	M3 & 10.91\% & 3.38\%  & 2.29\%   & M3 & 4.65\%  & 2.60\%  & 2.47\%   \\
	M4 & 3.57\%  & 1.99\%  & 1.87\%   & M4 & 1.89\%  & 1.22\%  & 1.10\%   \\
	M5 & 2.03\%  & 1.32\%  & 1.14\%   & M5 & 0.58\%  & 0.45\%  & 0.41\%   \\ \hline
	\multicolumn{4}{c|}{BULL2}               & \multicolumn{4}{c}{BEAR2}              \\ \hline
	Module   & Within  & Flow in & Flow out & Module   & Within  & Flow in & Flow out \\
	Name     & Modules & Modules & Modules  & Name     & Modules & Modules & Modules  \\ \hline
	M1 & 71.35\% & 9.17\%  & 10.60\%  & M1 & 82.16\% & 6.24\%  & 7.36\%   \\
	M2 & 11.49\% & 4.95\%  & 3.57\%   & M2 & 6.92\%  & 2.90\%  & 3.11\%   \\
	M3 & 6.82\%  & 2.85\%  & 3.08\%   & M3 & 3.91\%  & 2.40\%  & 1.83\%   \\
	M4 & 5.04\%  & 2.43\%  & 2.56\%   & M4 & 2.13\%  & 1.40\%  & 1.14\%   \\
	M5 & 1.92\%  & 1.11\%  & 1.23\%   & M5 & 2.01\%  & 1.29\%  & 1.05\%   \\ \hline
	\multicolumn{4}{c|}{BULL3}               & \multicolumn{4}{c}{BEAR3}              \\ \hline
	Module   & Within  & Flow in & Flow out & Module   & Within  & Flow in & Flow out \\
	Name     & Modules & Modules & Modules  & Name     & Modules & Modules & Modules  \\ \hline
	M1 & 8.34\%  & 5.05\%  & 5.08\%   & M1 & 46.54\% & 11.75\% & 11.49\%  \\
	M2 & 6.58\%  & 3.81\%  & 3.95\%   & M2 & 18.76\% & 8.26\%  & 8.20\%   \\
	M3 & 4.59\%  & 3.18\%  & 3.30\%   & M3 & 10.13\% & 5.21\%  & 5.08\%   \\
	M4 & 4.41\%  & 2.85\%  & 2.85\%   & M4 & 8.70\%  & 4.27\%  & 4.31\%   \\
	M5 & 3.15\%  & 1.87\%  & 1.98\%   & M5 & 7.09\%  & 3.17\%  & 3.43\%   \\ \hline
	\multicolumn{4}{c|}{BULL4}               & \multicolumn{4}{c}{BEAR4}              \\ \hline
	Module   & Within  & Flow in & Flow out & Module   & Within  & Flow in & Flow out \\
	Name     & Modules & Modules & Modules  & Name     & Modules & Modules & Modules  \\ \hline
	M1 & 7.45\%  & 4.30\%  & 4.43\%   & M1 & 13.18\% & 7.55\%  & 8.91\%   \\
	M2 & 6.46\%  & 2.60\%  & 2.17\%   & M2 & 12.64\% & 4.35\%  & 3.85\%   \\
	M3 & 5.49\%  & 3.72\%  & 3.71\%   & M3 & 7.57\%  & 3.72\%  & 3.83\%   \\
	M4 & 4.47\%  & 2.45\%  & 2.63\%   & M4 & 4.91\%  & 3.02\%  & 2.58\%   \\
	M5 & 4.28\%  & 2.71\%  & 2.77\%   & M5 & 4.53\%  & 2.21\%  & 2.05\%   \\ \hline
\end{tabular}
	
\end{table}

To investigate how modules develop and evolve over the eight periods, we consider stock categories in different modules and present results of BEAR1, BULL4, and BEAR4 in Tables~\ref{Tab Industry Categories of Stocks in BEAR1} to ~\ref{Tab Industry Categories of Stocks in BEAR4}, respectively.

\begin{table}[H]
	\centering
	\caption{Industry categories of stocks in top six modules during BEAR1.}
	\label{Tab Industry Categories of Stocks in BEAR1}	
	
	\begin{tabular}{lcccccc}
		\hline
		Module name & M1 & M2 & M3 & M4 & M5 & M6 \\ \hline
		Materials                   & 102      & 6        & 0        & 6        & 1        & 0        \\
		Telecommunication service & 2        & 0        & 0        & 0        & 0        & 0        \\
		Real estate                & 27       & 6        & 21       & 3        & 1        & 0        \\
		Industrials                   & 146      & 7        & 7        & 4        & 1        & 0        \\
		Utilities           & 38       & 0        & 1        & 2        & 0        & 0        \\
		Finance                  & 13       & 1        & 2        & 0        & 0        & 4        \\
		Consumer staples       & 107      & 12       & 1        & 1        & 0        & 0        \\
		Energy                     & 20       & 1        & 1        & 1        & 1        & 0        \\
		Consumer discretionary          & 54       & 2        & 0        & 0        & 0        & 0        \\
		Information technology     & 45       & 4        & 0        & 2        & 0        & 0        \\
		Health care               & 55       & 5        & 0        & 0        & 0        & 0        \\ \hline
	\end{tabular}
	
\end{table}

As shown in Table~\ref{Tab Industry Categories of Stocks in BEAR1}, the SSE A-shares network in the BEAR1 can be divided into nine modules, and three modules are out of discussion due to only including one stock. M1 has the largest size ($ 608 $ stocks) in the network, includes, and captures $ 85.87\% $ of the network's information. Conversely, the remaining modules contain less information, and the differentiation effect caused by industry aggregation is not significant. More importantly, the industry distribution of the M1 in the BEAR1 is analogous to that of the entire SSE A-shares market distribution. Consequently, M1 can be viewed as a system module, and other stocks that are not in the first module are peripheral in the stock market. Similar patterns can also be found in the other first four stages.

Unlike the first four stages, the last four stages except for BEAR3 present the significant industry differentiation of the modules, as shown in Tables~\ref{Tab Industry Categories of Stocks in BULL4} and \ref{Tab Industry Categories of Stocks in BEAR4}.

\begin{table}[H]
	\centering
	\caption{Industry categories of stocks in top nine modules during BULL4.}
	\label{Tab Industry Categories of Stocks in BULL4}
	\footnotesize
	\begin{tabular}{lccccccccc}
		\hline
		Module name & M1 & M2 & M3 & M4 & M5 & M6 & M7 & M8 & M9 \\ \hline
		Materials               & 2        & 1        & 8        & 2        & 1        & 3        & 2        & 25       & 1        \\
		Real estate            & 1        & 0        & 1        & 1        & 4        & 1        & 2        & 0        & 29       \\
		Industrials               & 1        & 2        & 5        & 21       & 14       & 7        & 7        & 2        & 1        \\
		Utilities       & 2        & 1        & 8        & 6        & 4        & 1        & 0        & 0        & 4        \\
		Finance              & 0        & 0        & 1        & 6        & 0        & 2        & 1        & 0        & 0        \\
		Consumer staples   & 5        & 4        & 2        & 1        & 1        & 3        & 11       & 5        & 3        \\
		Energy                 & 0        & 0        & 0        & 0        & 0        & 0        & 0        & 0        & 0        \\
		Consumer discretionary      & 0        & 14       & 4        & 1        & 1        & 2        & 2        & 4        & 0        \\
		Information technology & 1        & 3        & 1        & 0        & 2        & 10       & 0        & 0        & 0        \\
		Health care           & 28       & 8        & 0        & 1        & 2        & 1        & 2        & 0        & 0        \\ \hline
	\end{tabular}
	
\end{table}

Table~\ref{Tab Industry Categories of Stocks in BULL4} suggests that and nine largest modules in the BULL4 stage network and the rest $ 42 $ modules are out of discussion due to involving few stocks. The differentiation of the modules displays a significant industry aggregation phenomenon. Health care is the dominant industry category in M1 and is key to the information flow and spillover risk. Given by Figure~\ref{Fig Module Visualisation}, the closed connections between M1 and M2 implies the high potential risk of contagion between the health care and the consumer discretionary industry because of the distinct overlap of these two industries in the industry chain.

\begin{table}[H]
	\centering
	\caption{Industry categories of stocks in top nine modules during BEAR4.}
	\label{Tab Industry Categories of Stocks in BEAR4}	
	
	\begin{tabular}{lccccccccc}
		\hline
		Module name & M1 & M2 & M3 & M4 & M5 & M6 & M7 & M8 & M9 \\ \hline
		Materials                   & 15       & 3        & 1        & 1        & 1        & 18       & 5        & 0        & 1        \\
		Telecommunication service & 0        & 0        & 0        & 1        & 0        & 0        & 0        & 0        & 0        \\
		Real estate                & 9        & 0        & 1        & 2        & 1        & 5        & 9        & 0        & 2        \\
		Industrials                   & 22       & 4        & 19       & 16       & 4        & 3        & 7        & 4        & 7        \\
		Utilities           & 8        & 0        & 0        & 0        & 1        & 0        & 6        & 1        & 0        \\
		Finance                  & 1        & 0        & 0        & 0        & 0        & 4        & 0        & 1        & 0        \\
		Consumer staples       & 20       & 10       & 3        & 1        & 18       & 2        & 1        & 10       & 0        \\
		Energy                     & 3        & 0        & 0        & 0        & 1        & 8        & 0        & 0        & 0        \\
		Consumer discretionary          & 3        & 15       & 1        & 0        & 5        & 0        & 1        & 0        & 2        \\
		Information technology     & 7        & 0        & 27       & 1        & 2        & 1        & 1        & 1        & 0        \\ \hline
	\end{tabular}	
	
\end{table}

The BEAR4 stage can be divided into $ 46 $ modules, and the top nine modules with the largest sizes capture $ 57.44\% $ of the network information. The industry categories of stocks in these nine modules are provided in Table~\ref{Tab Industry Categories of Stocks in BEAR4}. M1 mainly consists of stocks from materials, industry, and consumer staples. The dominant industries in M2, M3, M4, M5, and M6 are consumer discretionary and consumer staples, information technology and industrials, industrials, consumer staples, and materials.

Tables~\ref{Tab Industry Categories of Stocks in BULL4} and \ref{Tab Industry Categories of Stocks in BEAR4} demonstrate the significant industry aggregation effect in the BULL4 and BEAR4 period. With the development of the stock market, the industry aggregation results in module differentiation in SSE A-shares because stocks from the same industry category share similar macro fundamentals.

In summary, more than $ 70\% $ of stocks in the SSE A-shares market belong to the same module in the first four periods, which can be viewed as the system module. Stocks in the system module are less connected with others, and those outside the system module are pericardial in the network, implying the relative low contagion risk and weak connections. As China's financial market develops, the system module is gradually differentiated into several small parts based on industry categories. Stocks from the same or related industries grouping in modules make the industry aggregation phenomenon more significant. In an extreme situation like the BEAR3 stage, abnormally fluctuations and large-scale declines in the market lead to the formation of a small system module whose higher status accelerates risk contagion.


\subsection{Topological Properties of SSE A-Shares Networks}

We utilize topological indicators in Section \ref{Section Network topological indicators} to investigate the SSE A-shares market's characteristics over bull and bear markets. Details are presented in Tables~\ref{Tab Characteristics of SSE A-shares Networks First Half Period} and \ref{Tab Characteristics of SSE A-shares Networks Second Half Period}.

\begin{table}[H]
	\centering
	\caption{Characteristics of the SSE A-shares networks during the first four stages.}
	\label{Tab Characteristics of SSE A-shares Networks First Half Period}
	
	\begin{tabular}{lcccc}
		\hline
		Topological properties                  & BULL1   & BEAR1   & BULL2   & BEAR2   \\ \hline
		Network diameter                        & 7       & 6       & 6       & 7       \\
		Network density                         & 0.0195  & 0.0252  & 0.02    & 0.0342  \\
		The average shortest path length                 & 2.72    & 2.7     & 2.85    & 2.59    \\
		Clustering coefficient                  & 0.071   & 0.078   & 0.055   & 0.097   \\
		Mean of relative degree centrality      & 0.02    & 0.0252  & 0.02    & 0.0342  \\
		Mean of relative betweenness centrality & 0.23    & 0.234   & 0.255   & 0.22    \\
		Mean of relative closeness centrality   & 0.3815  & 0.3909  & 0.3699  & 0.4138  \\
		Out-degree centralisation               & 0.11404 & 0.05463 & 0.04857 & 0.11844 \\
		In-degree centralisation                & 0.03841 & 0.13446 & 0.11719 & 0.06242 \\
		Betweenness centralisation              & 0.012   & 0.0167  & 0.0308  & 0.0162  \\
		Out-degree closeness centralisation     & 0.2976  & 0.1937  & 0.1891  & 0.275   \\
		In-degree closeness centralisation      & 0.1742  & 0.345   & 0.336   & 0.1831  \\ \hline
	\end{tabular}	
	
\end{table}

\begin{table}[H]
	\centering
	\caption{Characteristics of the SSE A-shares networks during the last four stages.}
	\label{Tab Characteristics of SSE A-shares Networks Second Half Period}
	
	\begin{tabular}{lcccc}
		\hline
		Topological properties                  & BULL3   & BEAR3   & BULL4   & BEAR4   \\ \hline
		Network diameter                        & 8       & 6       & 6       & 8       \\
		Network density                         & 0.0165  & 0.022   & 0.0193  & 0.0157  \\
		The average shortest path length        & 3.1     & 2.84    & 3.08    & 3.23    \\
		Clustering coefficient                  & 0.066   & 0.076   & 0.084   & 0.075   \\
		Mean of relative degree centrality      & 0.0165  & 0.022   & 0.0193  & 0.0157  \\
		Mean of relative betweenness centrality & 0.28    & 0.255   & 0.278   & 0.296   \\
		Mean of relative closeness centrality   & 0.3326  & 0.3748  & 0.3389  & 0.3209  \\
		Out-degree centralisation               & 0.09127 & 0.01859 & 0.15156 & 0.07114 \\
		In-degree centralisation                & 0.04645 & 0.09201 & 0.06052 & 0.05434 \\
		Betweenness centralisation              & 0.0361  & 0.0172  & 0.0448  & 0.025   \\
		Out-degree closeness centralisation     & 0.2952  & 0.083   & 0.4127  & 0.2957  \\
		In-degree closeness centralisation      & 0.2062  & 0.2378  & 0.2306  & 0.2169  \\ \hline
	\end{tabular}		
	
\end{table}

As shown in Tables~\ref{Tab Characteristics of SSE A-shares Networks First Half Period} and \ref{Tab Characteristics of SSE A-shares Networks Second Half Period}, network densities of bear markets are higher than those of bull markets, and network diameters of bear markets are lower than those of bull markets, suggesting that stocks have stronger connections with others in the bear market. Affected by the subprime mortgage crisis and the deleveraging of capital allocation, the SSE A-shares market experiences large-scale collapses and shows the significant connected effect. Meanwhile, investors are more sensitive to market information in these two periods, and hence, tend to adopt similar strategies to avoid risks.

In the entire period, the lengths of the average shortest path in the SSE A-shares networks are between two and three, meaning that approximately two or three intermediate stocks can connect any stock pairs. The clustering coefficients of bear markets are generally higher than those of bull markets, reflecting the more distinct connected effect in recessions. The BEAR1 is during the subprime mortgage crisis, the BEAR2 is affected by the European debt crisis, and the BEAR3 experiences the ``thousand-share limit-down" after the stock market deleverages in 2015, which reflects a financial crisis enhances the small-world effect of the SSE A-shares network.

Regarding the difference between bull and bear markets, the average degrees of stocks in bear markets are greater than those in bull markets. In the bear market, the core stocks have more leading influence, and more synchronous changes appear in the market. Also, closeness centralities are relatively high, which means that the reachable distances of risk propagation are relatively short, and risks can transmit to most stocks from the source via a short distance.

From the structural differences between bear markets, BEAR1 and BEAR3 show rapid declines with large volatility, while BEAR2 and BEAR4 fall with fluctuations and have small volatility. Therefore, structural differences exist between these two types of bear markets. The low out-degree centralizations of BEAR1 and BEAR3 reflect the marginal differences between the out-degree centrality of each node and the maximum out-degree centrality. Therefore, most stocks in these two periods have relatively high out degrees, and risks are more likely to transmit to other stocks through these stronger connections, accelerating the propagation of risks. By contrast, the BEAR2 and BEAR4 markets show different patterns. The low in-degree centralization and high out-degree centralizations reflect that most stocks have relatively high in degrees but low out degrees, suggesting that stocks in the network absorb risks and prevent the spread of risks. BEAR1 and BEAR3 are periods with abnormal fluctuations due to the impact of the subprime mortgage crisis and rapid deleveraging in 2015. In BEAR1 and BEAR2, it is hard to identify critical stocks leading to the massive collapse, making the supervision on risks more challenging. Conversely, the less connected structure of the SSE A-shares networks in BEAR2 and BEAR4 lowers the transmission risks and benefits of identifying the risk source and restraining the network's spread of risk.


\subsection{Analysis of Core Stocks in Bull and Bear Markets}

We use three types of centralities to measure stocks' influence in the SSE A-shares market to identify core stocks in the network and investigate how these stocks transmit risks over different periods. Tables~\ref{Tab Top Five Relative Degree Centralities} to \ref{Tab Top Five Relative Closeness Centralities} list stocks with the top five relative degrees, betweenness and closeness centralities in eight stages, respectively, where the first column reports stock codes in the SSE A-share market and the second column gives the corresponding industry categories. Stocks in these tables are referred to as core stocks due to their distinct influence in specific periods.

\begin{table}[H]
	\centering
	\caption{Stocks with Top Five Relative \textbf{Degree} Centralities in Eight Periods.}
	\label{Tab Top Five Relative Degree Centralities}
	
	\begin{tabular}{cl|cl}
		\hline
		\multicolumn{2}{c|}{BULL1}      & \multicolumn{2}{c}{BEAR1}      \\ \hline
		600624 & Health care            & 600340 & Real estate            \\
		600218 & Industrials            & 600085 & Health care            \\
		600373 & Consumer discretionary & 600088 & Consumer discretionary \\
		600172 & Materials              & 600006 & Consumer discretionary \\
		600626 & Consumer discretionary & 600590 & Industrials            \\ \hline
		\multicolumn{2}{c|}{BULL2}      & \multicolumn{2}{c}{BEAR2}      \\ \hline
		600405 & Industrials            & 600370 & Consumer discretionary \\
		600967 & Industrials            & 600853 & Industrials            \\
		600665 & Real estate            & 600522 & Information technology \\
		600692 & Real estate            & 600590 & Industrials            \\
		600601 & Industrials            & 600360 & Information technology \\ \hline
		\multicolumn{2}{c|}{BULL3}      & \multicolumn{2}{c}{BEAR3}      \\ \hline
		600460 & Information technology & 600410 & Information technology \\
		600439 & Consumer staples       & 600502 & Industrials            \\
		660360 & Information technology & 600271 & Information technology \\
		600131 & Utilities              & 600749 & Consumer discretionary \\
		600345 & Information technology & 600531 & Materials              \\ \hline
		\multicolumn{2}{c|}{BULL4}      & \multicolumn{2}{c}{BEAR4}      \\ \hline
		600229 & Consumer discretionary & 600168 & Utilities              \\
		600561 & Industrials            & 600331 & Materials              \\
		600356 & Materials              & 600292 & Industrials            \\
		600757 & Consumer discretionary & 600713 & Health care            \\
		600422 & Health care            & 600269 & Industrials            \\ \hline
	\end{tabular}
	
\end{table}

As suggested in Tables~\ref{Tab Top Five Relative Degree Centralities} to \ref{Tab Top Five Relative Closeness Centralities}, high influence stocks are mainly from four industry categories: consumer discretionary, health care, materials, and industrials, which corresponds to the system module and the industry aggregation phenomenon demonstrated in Tables~\ref{Tab Industry Categories of Stocks in BEAR1} to \ref{Tab Industry Categories of Stocks in BEAR4}. Moreover, there exists a great joint part in three stock lists measured by different centralities. A few stocks are even included by three lists in the same stage, implying these stocks play irreplaceable roles in the system during the particular time. Also, bull and bear markets have noticeably different core stocks. About $ 22.5\% $ of stocks are ranked in the top 20 stocks regarding one type of centralities in a specific stage, meaning the diversity of core stocks in eight stages and the complicated leading effect raised by these stocks.

\begin{table}[H]
	\centering
	\caption{Stocks with Top Five Relative \textbf{Betweenness} Centralities in Eight Periods.}
	\label{Tab Top Five Relative Betweenness Centralities}
	
	\begin{tabular}{cl|cl}
		\hline
		\multicolumn{2}{c|}{BULL1}      & \multicolumn{2}{c}{BEAR1}      \\ \hline
		600373 & Consumer discretionary & 600798 & Industrials            \\
		600353 & Information technology & 600088 & Consumer discretionary \\
		600624 & Health care            & 600811 & Consumer staples       \\
		600138 & Consumer discretionary & 600736 & Real estate            \\
		600426 & Materials              & 600565 & Real estate            \\ \hline
		\multicolumn{2}{c|}{BULL2}      & \multicolumn{2}{c}{BEAR2}      \\ \hline
		600692 & Real estate            & 600480 & Consumer discretionary \\
		600967 & Industrials            & 600131 & Utilities              \\
		600665 & Real estate            & 600370 & Consumer discretionary \\
		600229 & Consumer discretionary & 600585 & Materials              \\
		600662 & Industrials            & 600004 & Industrials            \\ \hline
		\multicolumn{2}{c|}{BULL3}      & \multicolumn{2}{c}{BEAR3}      \\ \hline
		600131 & Utilities              & 600719 & Utilities              \\
		600439 & Consumer staples       & 600410 & Information technology \\
		600460 & Information technology & 600533 & Real estate            \\
		600166 & Consumer discretionary & 600502 & Industrials            \\
		600879 & Industrials            & 600501 & Industrials            \\ \hline
		\multicolumn{2}{c|}{BULL4}      & \multicolumn{2}{c}{BEAR4}      \\ \hline
		600135 & Materials              & 600331 & Materials              \\
		600561 & Industrials            & 600594 & Health care            \\
		600422 & Health care            & 600713 & Health care            \\
		600757 & Consumer discretionary & 600375 & Industrials            \\
		600730 & Consumer discretionary & 600390 & Finance                \\ \hline
	\end{tabular}
	
\end{table}

\begin{table}[H]
	\centering
	\caption{Stocks with Top Five Relative \textbf{Closeness} Centralities in Eight Periods.}
	\label{Tab Top Five Relative Closeness Centralities}
	
	\begin{tabular}{cl|cl}
		\hline
		\multicolumn{2}{c|}{BULL1}      & \multicolumn{2}{c}{BEAR1}       \\ \hline
		600624 & Health care            & 600085 & Health care            \\
		600172 & Materials              & 600790 & Real estate            \\
		600983 & Consumer discretionary & 600006 & Consumer discretionary \\
		600373 & Consumer discretionary & 600811 & Consumer staples       \\
		600985 & Energy                 & 600590 & Industrials            \\ \hline
		\multicolumn{2}{c|}{BULL2}      & \multicolumn{2}{c}{BEAR2}       \\ \hline
		600405 & Industrials            & 600370 & Consumer discretionary \\
		600967 & Industrials            & 600522 & Information technology \\
		600692 & Real estate            & 600853 & Industrials            \\
		600665 & Real estate            & 600480 & Consumer discretionary \\
		600787 & Industrials            & 600861 & Consumer staples       \\ \hline
		\multicolumn{2}{c|}{BULL3}      & \multicolumn{2}{c}{BEAR3}       \\ \hline
		600439 & Consumer staples       & 600410 & Information technology \\
		600460 & Information technology & 600749 & Consumer discretionary \\
		600131 & Utilities              & 600502 & Industrials            \\
		600345 & Information technology & 600271 & Information technology \\
		600166 & Consumer discretionary & 600661 & Consumer discretionary \\ \hline
		\multicolumn{2}{c|}{BULL4}      & \multicolumn{2}{c}{BEAR4}       \\ \hline
		600229 & Consumer discretionary & 600331 & Materials              \\
		600561 & Industrials            & 600168 & Utilities              \\
		600218 & Industrials            & 600713 & Health care            \\
		600757 & Consumer discretionary & 600292 & Industrials            \\
		600422 & Health care            & 600320 & Industrials            \\ \hline
	\end{tabular}
	
\end{table}

We present the averages of financial indicators of stocks with the top 20 centralities in different stages in Table~\ref{Tab Financial Features of Top 20 Stocks} to further investigate characteristics of core stocks. For comparisons, Table~\ref{Tab Financial Features of Top 20 Stocks} also provides the financial indicator averages of all SSE A-shares in parentheses.

\begin{table}[H]
	\centering
	\caption{Financial characteristics of stocks in the SSE A-shares network over eight periods. Numbers in cells and parentheses report the financial indicator averages of stocks with the top 20 centralities and all SSE A-shares.}
	\label{Tab Financial Features of Top 20 Stocks}
	
	\begin{tabular}{crrrrrr}
		\hline
		Period & \multicolumn{1}{c}{Current} & \multicolumn{1}{c}{Quick} & \multicolumn{1}{c}{Debt to asset}     & \multicolumn{1}{c}{Total assets}   & \multicolumn{1}{c}{Return on} & \multicolumn{1}{c}{Margin trading} \\
		& \multicolumn{1}{c}{ratio}   & \multicolumn{1}{c}{ratio}        & \multicolumn{1}{c}{ratio} & \multicolumn{1}{c}{turnover ratio} & \multicolumn{1}{c}{equity}    & \multicolumn{1}{c}{balance}        \\ \hline
		BULL1  & 1.42 (1.38)                 & 1.02 (1.43)                      & 59.33 (43.03)                  & 0.69 (0.81)                         & 15.26 (6.05)            & - (-)                              \\
		BEAR1  & 1.42 (1.22)                 & 0.98 (0.71)                      & 60.09 (55.94)                  & 0.78 (0.78)                         & 11 (11.23)              & - (-)                              \\
		BULL2  & 1.45 (1.55)                 & 0.92 (0.85)                      & 62.47 (44.29)                  & 0.79 (0.86)                         & 5.30 (7.40)             & - (-)                              \\
		BEAR2  & 1.71 (1.08)                 & 1.14 (1.51)                      & 58.95 (47.17)                  & 0.79 (0.60)                         & 10.18 (3.30)            & 1.25 (0.48)                        \\
		BULL3  & 1.65 (1.81)                 & 1.26 (1.03)                      & 54.93 (43.48)                  & 0.77 (0.57)                         & 10.86 (5.98)            & 5.14 (2.31)                        \\
		BEAR3  & 1.94 (1.83)                 & 1.48 (1.41)                      & 49.28 (55.78)                  & 0.71 (0.54)                         & -0.07 (-4.09)           & 6.52 (6.64)                        \\
		BULL4  & 1.95 (2.54)                 & 1.48 (2.16)                      & 49.84 (43.20)                  & 0.64 (0.71)                         & 6.82 (4.57)             & 4.76 (4.00)                        \\
		BEAR4  & 1.95 (1.66)                 & 1.50 (1.28)                      & 50.13 (48.31)                  & 0.66 (0.76)                         & -11.39 (-3.98)          & 4.34 (3.08)                        \\ \hline
	\end{tabular}
	
\end{table}

Table~\ref{Tab Financial Features of Top 20 Stocks} suggests that most core stocks (i.e., stocks with the top 20 centralities) are medium- or small-cap stocks rather than large-cap ones. Moreover, core stocks lack short-term solvency because about two-thirds of them have lower current ratios and quick ratios than the average levels of all SSE A-shares at the same stage, and the current ratio of nearly one-third of core stocks is below one. Regarding profitability, The return on net assets of core stocks declines annually and is far lower than the average value of SSE A-shares in the same period. The return on equity of two-thirds of core stocks is lower than $ 10\% $, reflecting low profitability.

In sum, core stocks that have the leading effect in the SSE A-shares network are mainly issued by medium- or small-cap companies with relatively low financial indicators. Conversely, the large-cap stocks weigh more in the stock index but are less connected with other stocks. and the herding behavior can explain this phenomenon in the stock market. In the bull market, individual investors, accounting for a significant proportion in the Chinese stock market, tend to follow the upward trend via buying medium or small-cap stocks. Compared with large-cap stocks, these stocks have relatively high growth rates in the short term because they have relatively small market values and are easily influenced by capitals. However, these stocks' rise does not derive from fundamentals like technological breakthroughs but from short-term arbitrage strategies, which is difficult to attract long-term investors' attention. Consequently, such rapid increases caused by blindness only quickly accumulate bubbles in a short period. Selling off these stocks at high prices, speculators also release a signal that stimulates individual investors to sell stocks with panic, which eventually leads to a steep decline in stock prices. Besides, the immature capital market and irrational behaviors widely spreading in the market result in widespread risks and the phenomenon of a sharp rise and fall for the entire market.

%


\section{Discussion}

Based on the daily closing prices of SSE A-shares from 2005 to 2018, this paper utilizes the minimum entropy method and topological properties of networks to investigate the evolution of the SSE A-shares market from macro and micro perspectives. The main research conclusions are listed as follows:

First, stocks in the SSE A-shares market are closely connected over all periods, and the connected effect is more significant in bear markets. As a result, the degree of declines in bear markets is much greater than that of rises in bull markets. 
No apparent risk sources exist in the rapidly falling bear markets, but those sources can be identified in the slowly falling bear markets, which is beneficial to control.

Second, the SSE A-shares network shows the industry differentiation in the last four stages. In the first four stages, most stocks belong to the same module, referred to as the system module, which implies that risk contagion mainly appears in this module and risks transmit from the central module to peripheral modules. As the Chinese financial market develops, the growing industry aggregation in the SSE A-shares market gives rise to the module differentiation and gradually undermines the system module's status. Consequently, risks are more likely to spread in modules with similar industry categories.

Third, some stocks have consistent influences on the SSE A-shares market over eight periods, and most of them belong to health care, consumer discretionary, industrials, and materials. Status analyses suggest that a few stocks have leading effects on others and play irreplaceable roles in the network. Also, medium- and small-cap stocks with poor financial conditions are more likely to become risk sources in the SSE A-shares network, especially in the bear stage.

The Chinese financial system's development increases investors' risk awareness, the number of institutional investors, fundamental analysis abilities, and the sensitivity to industry policies. As a result, the SSE A-shares market is expected to increase systematic differentiation and industry aggregation. This paper helps investors better understand this phenomenon and improve their portfolio strategies.

\bigskip

\clearpage

\bibliographystyle{apalike}
\bibliography{ReferenceMinimumInfoEntropy}

\clearpage

\appendix
\setcounter{table}{0}
\renewcommand{\thetable}{\Alph{table}}
\renewcommand{\thefigure}{\Alph{figure}}
\addtocontents{lot}{\bigskip}
\addtocontents{lot}{Appendix}

\appendix\label{Appendix A}
\section{Appendix}

\begin{figure}[H]

\includegraphics[width = \columnwidth]{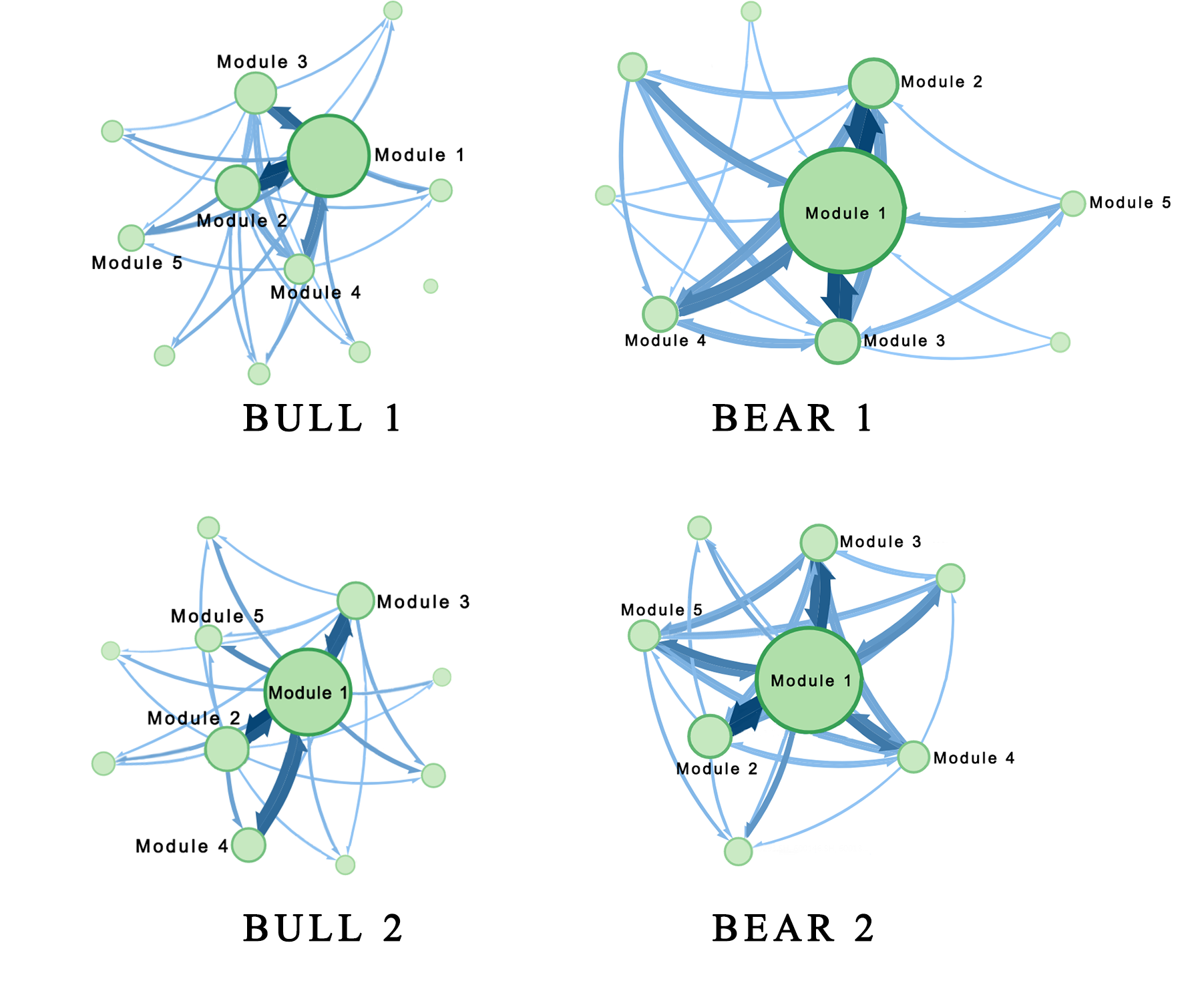}
\caption{Module divisions in eight periods based on the Map Equation algorithm. Nodes represent modules, directed links indicate directions of the information flow, and the thickness of links demonstrates the transmission probability of risk between different modules. The node size is proportional to the information flow in a module that includes the information flow out of and within the module. Specifically, the boundary thickness of a node is proportional to the information flow out of the module, corresponding to the probability that risks transmit to other modules. The interior area of a node is proportional to the information flow within the module, corresponding to the probability that risks staying in the module.}
\label{Fig Full Size 1}
	
\end{figure}

\begin{figure}[H]

\includegraphics[width = \columnwidth]{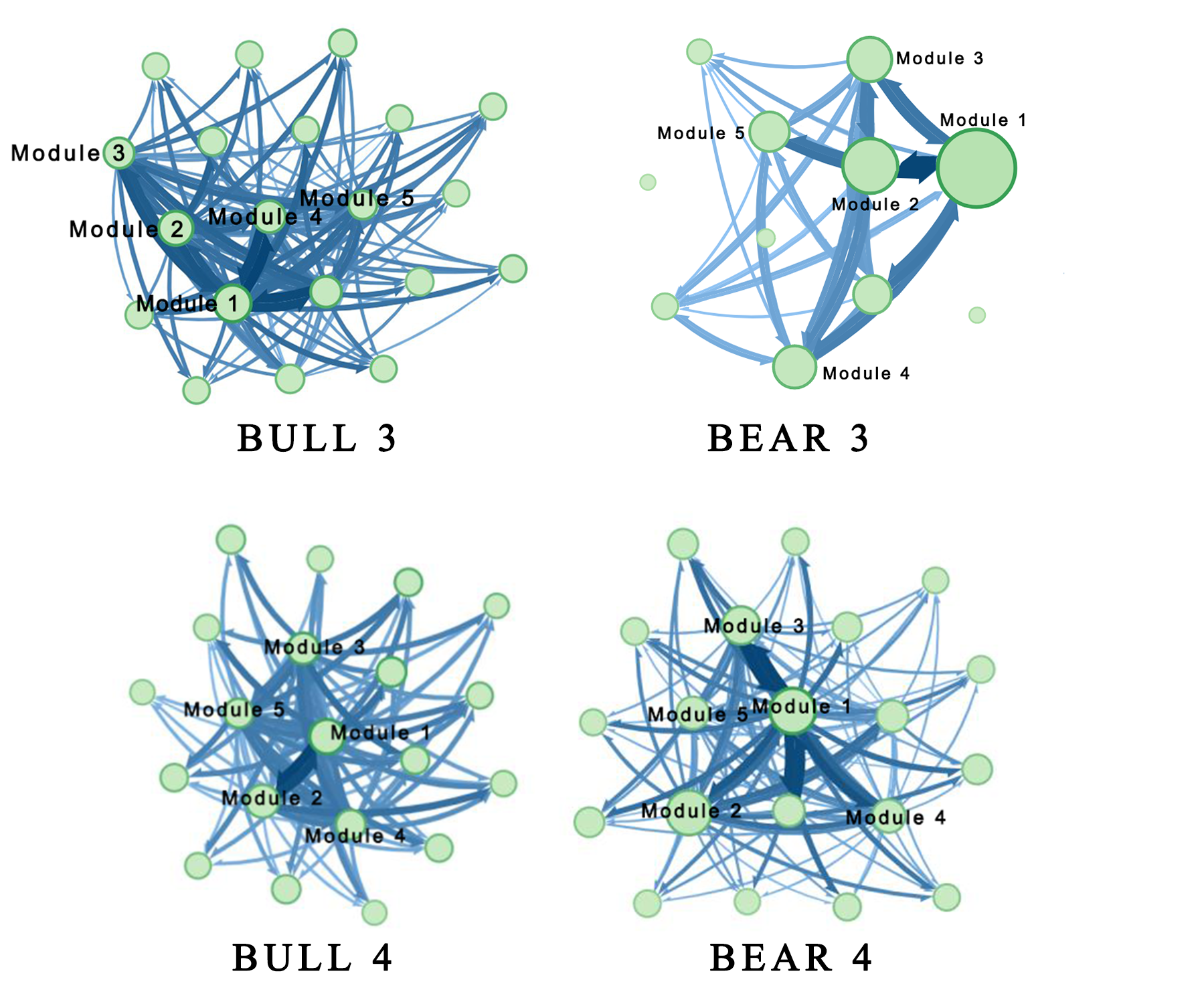}
\caption{Module divisions in eight periods based on the Map Equation algorithm. Nodes represent modules, directed links indicate directions of the information flow, and the thickness of links demonstrates the transmission probability of risk between different modules. The node size is proportional to the information flow in a module that includes the information flow out of and within the module. Specifically, the boundary thickness of a node is proportional to the information flow out of the module, corresponding to the probability that risks transmit to other modules. The interior area of a node is proportional to the information flow within the module, corresponding to the probability that risks staying in the module.}
\label{Fig Full Size 2}

\end{figure}

\end{document}